# Shubnikov–de Haas oscillations

# in *p* and *n*-type topological insulator (Bi$_x$Sb$_{1-x}$)$_2$Te$_3$

Ryota Akiyama[1][1], Kazuki Sumida[2], Satoru Ichinokura[1], Ryosuke Nakanishi[1], Akio Kimura[2], Konstantin A. Kokh[3,4,5], Oleg E. Tereshchenko[3,4,6], and Shuji Hasegawa[1]

[1]*Department of physics, The University of Tokyo, 7-3-1 Hongo, Bunkyo-ku, Tokyo 113-0033, Japan*

[2]*Department of Physics, Hiroshima University, 1-3-1 Kagamiyama, Higashi-Hiroshima, Hiroshima 739-8526, Japan*

[3]*Novosibirsk State University, ul. Pirogova 2, Novosibirsk, 630090, Russia*

[4]*Saint Petersburg State University, Saint Petersburg 198504, Russia*

[5] *V.S. Sobolev Institute of Geology and Mineralogy, Novosibirsk 630090, Russia*

[6] *A.V. Rzhanov Institute of Semiconductor Physics, Novosibirsk 630090, Russia*

**Abstract**

We show Shubnikov-de Haas (SdH) oscillations in topological insulator (Bi$_x$Sb$_{1-x}$)$_2$Te$_3$ flakes whose carrier types are *p*-type ($x$ = 0.29, 0.34) and *n*-type ($x$ = 0.42). The physical properties such as the Berry phase, carrier mobility, and scattering time significantly changed by tuning the Fermi-level position with the concentration $x$. The analyses of SdH oscillations by Landau-level fan diagram, Lifshitz−Kosevich theory, and Dingle-plot in the $p$-type samples with $x$ = 0.29 and 0.34 showed the Berry phase of zero and a relatively low mobility (2000 – 6000 cm$^2$/V/s). This is due to the dominant bulk component in transport. On the other hand, the mobility in the $n$-type sample with $x$ = 0.42 reached a very large value ~ 17,000 cm$^2$/V/s and the Berry phase of near π, whereas the SdH oscillations were neither purely two- nor three-dimensional. These suggest that the transport channel has a surface-bulk coupling state which makes the carrier scattering lesser and enhances the mobility and has a character between two- and three-dimension.

---

[1] akiyama@surface.phys.s.u-tokyo.ac.jp

The topological surface states (TSSs) on topological insulators (TIs) [1,2] have attracted much attention recently because they are one of physical states induced by the intrinsically lifted spin degeneracy in the band structures together with spin-momentum locking relation. Thus, they have abundant potential for applications to spintronics because the TSSs do not need external magnetic field to lift the spin degeneracy and generate spin current as well as various kinds of spin-related phenomena. However, the bulk contribution in the electrical transport for real crystals has often interfered with the observation of the TSS-driven physical properties. To solve this problem, the tuning of the Fermi-level position has been tried by various methods such as electrical gating [3-6] and chemical doping of carriers [7-10].

On the other hand, as one of the most promising TIs, a ternary alloy $(Bi_xSb_{1-x})_2Te_3$ is proposed, which can be both $p$- and $n$-type by tuning the Fermi level position with controlling the concentration $x$ [11-15], since the binary sesquichalcogenides $Bi_2Te_3$ and $Sb_2Te_3$ are degenerate $n$- and $p$-type for real crystals, respectively. The tuning of the Fermi level position can make the bulk insulating, resulting in reduction of the bulk carrier concentration. In Ref. [13], Zhang *et al.* measured band structures of $(Bi_xSb_{1-x})_2Te_3$ films with $x$ = 0, 0.25, 0.62, 0.75, 0.88, 0.94, 0.96 and 1.0 by angle-resolved photoemission spectroscopy (ARPES). They insist that they could see the surface band in all samples along Γ-M and Γ-K direction in the range of $k \leq 0.2$ Å$^{-1}$. On the other hand, Kong *et al.* showed the ambipolar control of the carrier type in (Bi$_x$Sb$_{1-x}$)$_2$Te$_3$ by changing gate voltage in appropriate sample with $x = 0.5$ [11]. By changing both $x$ and the gate voltage, the transport characteristics such as the temperature dependence of resistivity can be tuned. In addition, they checked samples by ARPES also. However, because the Shubnikov-de Haas oscillation has not been reported, the specific physical parameters such as 2D carrier density, cyclotron mass, scattering time, mean free path, Fermi wave number, mobility, and the Berry phase have not been clarified in (Bi$_x$Sb$_{1-x}$)$_2$Te$_3$. In Ref. [14], Yoshimi *et al.* observed the quantum Hall (QH) effect in the structure of (Bi$_x$Sb$_{1-x}$)$_2$Te$_3$/Cr-doped (Bi$_x$Sb$_{1-x}$)$_2$Te$_3$ with applying the gate voltage. They revealed that the surface state possesses $\nu = 0$ and $+1$ QH states under magnetic fields with dominant contribution from the anomalous Hall effect in magnetic topological insulator layer. The appearance of these QH states are explained by the magnetization-induced gap and/or the Landau Levels at the surface state. However, in these works fundamental physical parameters have not been estimated enough.

The TSS band dispersion of (Bi$_x$Sb$_{1-x}$)$_2$Te$_3$ was observed by ARPES and time-resolved ARPES (TARPES) [11,16]. As for electrical transport measurements, the quantum Hall effect was confirmed with electrical gating [14]. The doping of magnetic atoms such as Cr and V in $(Bi_xSb_{1-x})_2Te_3$ was also demonstrated for observing the quantum anomalous Hall effect, leading to possible spintronics devices [17,18].

Meanwhile, detailed systematic investigations of electronic properties in $(Bi_xSb_{1-x})_2Te_3$, such as the Berry phase, carrier mobility, and scattering time with changing $x$, have not yet been explored enough. In this paper, we investigated electrical transport properties of $(Bi_xSb_{1-x})_2Te_3$ samples with $x$ = 0.29, 0.34, and 0.42 (Samples A, B, and C, respectively, hereafter). The concentration $x$ was measured by Electron Probe Micro Analyzer (EPMA). We estimated the physical parameters by observing the Shubnikov-de Haas (SdH) oscillations. The transport in Samples A and B contained dominantly the bulk contribution. On the other hand, in Sample C, although the intercept of the Landau index in the Landau-level fan diagram was estimated to be -0.49 ± 0.09, which seemed to indicate the Berry phase of $\pi$ and dominance of the TSS transport, the transport channel consisted not only of the TSS component but also of the bulk one; the SdH oscillations showed neither two- nor three-dimensional characteristics. We can say that the transport channel in Sample C has a surface-bulk coupling state, which induces the lower scattering rate and the higher mobility (17,000 cm$^2$/V/s). Such nature is quite different from the cases of Samples A and B.

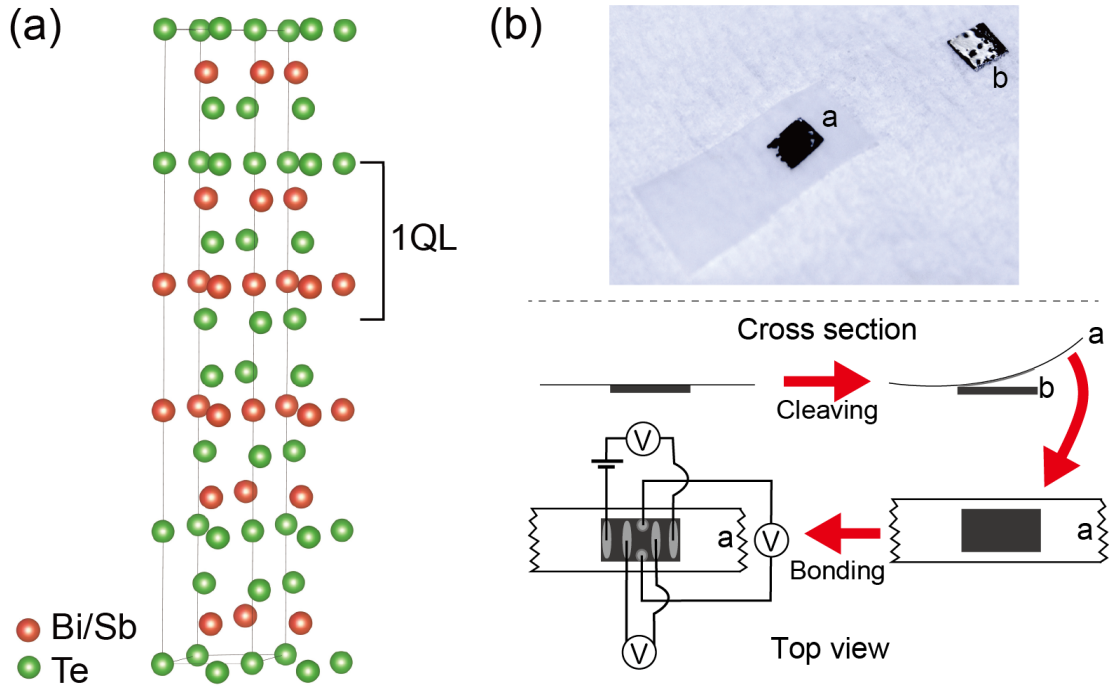

Fig. 1

FIG. 1 (Color online) (a) The crystal structure of $(Bi_xSb_{1-x})_2Te_3$. Green and red spheres represent Te and Bi/Sb atoms, respectively. The Bi and Sb atoms occupy the sites randomly. The quintuple layers of ~ 1 nm thick are bonded each other by Van der Waals force between Te layers. (b) A picture of the cleaved sample by adhesive tape (up) and a schematic picture of the preparation and the bonding of electrical wires (bottom).

Single crystals of $(Bi_xSb_{1-x})_2Te_3$ used in the present study were fabricated by the Bridgman method [19]. As shown in Fig. 1(a), the crystals have a layer structure in which five atomic layers (quintuple layer : QL) are one third of the lattice unit which are alternately stacked each other by Van der Waals bonds (between Te-Te layers). The samples were cleaved in air using an adhesive tape, and the flakes of 5 –30 μm-thick were formed on the tape; the thickness was confirmed by the 3D laser scanning microscope. The cleaved samples were bonded with Au wires by indium in air to make the configuration of the conventional six-probe technique as shown in Fig. 1(b). Electrical transport measurements were performed down to 2 K with magnetic fields perpendicular to the plane up to 14 T in a commercially available cryogenic equipment (PPMS by Quantum Design, Inc.).

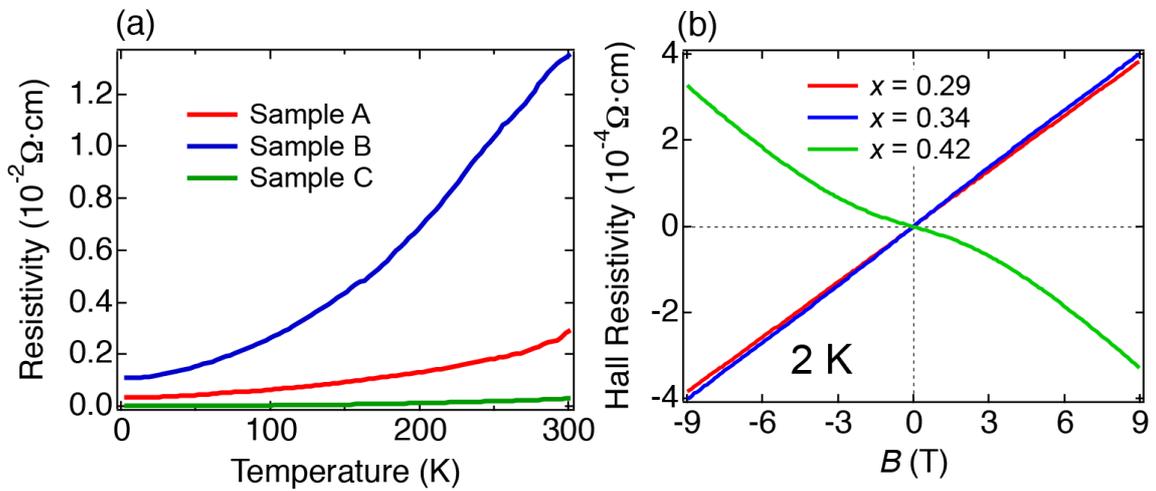

FIG. 2 (Color online) (a) The temperature dependences of the resistivity $\rho_{xx}$ down to 2 K. All samples (Samples A – C) show similar metallic characters. (b) The magnetic field dependence of the Hall resistivity at 2 K. The sign of the coefficient of the Hall resistivity in Sample C is different from the others, representing that the carrier type is $n$ ($p$) in Sample C (A and B).

Figure 2 (a) shows the temperature dependences of the resistivity $\rho_{xx}$ at 2 – 300 K for all samples. All curves represent the positive temperature coefficients, indicating a metallic conduction behavior. The resistivity of Sample C is significantly smaller than those of others, which is due to the high mobility as described later. For investigating the carrier type and density, the Hall measurement was done at 2 K as shown in Fig. 2 (b). From the coefficients of the Hall resistivity, Samples A and B were found to be *p*-type, whereas Sample C was *n*-type. Their carrier densities were estimated from the Hall coefficients within ± 3 T to be $2.2 \times 10^{19}$, $2.1 \times 10^{19}$, and $4.1 \times 10^{19}$ cm$^{-3}$, respectively.

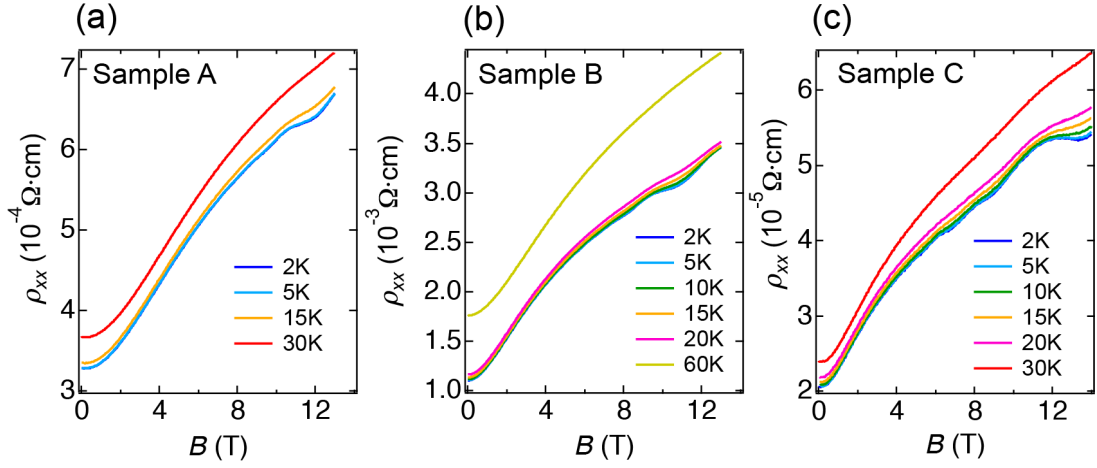

FIG. 3 (Color online) The magnetic field dependence of the longitudinal resistivity $\rho_{xx}$ up to 13 T ((a) Samples A and (b) B) or 14 T ((c) Sample C) at some selected temperatures within 2 – 30 K, respectively. The magnetic field was applied perpendicular to the sample surface. Clear Shubnikov-de Haas oscillations were seen at temperatures below ~ 15 K for all samples.

Figures 3 (a) - (c) show the magnetic-field dependence of the longitudinal resistivity $\rho_{xx}$ at some temperatures in the range of 2 – 60 K under the magnetic field up to 14 T. All samples show positive magnetoresistivity effect and also oscillatory changes at lower temperatures (< ~ 20 K) and higher magnetic fields (> ~ 5 T). For further analyzing such oscillations, inverse magnetic field dependences of $\rho_{xx}$, $\Delta\rho_{xx}$ and the fan diagram in Sample A, B, and C are shown in Figs. 4(a) - (c), respectively, where $\Delta\rho_{xx}$ is the resistivity deviation after subtracting the overall change in magnetoresistivity curves. Clear oscillations induced by the SdH effect can be seen for all samples in Fig. 4. The arrows indicate the positions of the corresponding oscillations. The appearance of SdH oscillations mean that Landau-levels pass through the Fermi level with increasing the magnetic field. The insets in $\Delta\rho_{xx}$ vs. $1/B$ graphs in Figs. 4(a) - (c) are the results of the Fast Fourier Transform (FFT) spectrum of the SdH oscillations at 2 K. Every FFT spectrum has a peak corresponding to the SdH oscillation, and the frequencies are 46, 36, and 26 T in Sample A, B, and C, respectively. The single prominent peak in the FFT spectra means that the channel is single or a surface-bulk coupled coherent transport for all samples [20-23].

One of the merits in observing the SdH oscillation in TIs is enabling us to know the Berry phase of the system [24-28]. By plotting the Landau index $n$ in the Landau-level fan diagram (the $\Delta\rho_{xx}$ peak as $n$ (filled square) and valley as $n+1/2$ (open square) as a function of $1/B$ ) as shown in the bottom figures in Fig. 4, we can estimate the intercept at $1/B = 0$ as in the insets of the fan diagram figures by fitting straight lines with the equation of $n = \alpha/B + \beta$ using Lifshitz-Onsager quantization rule.

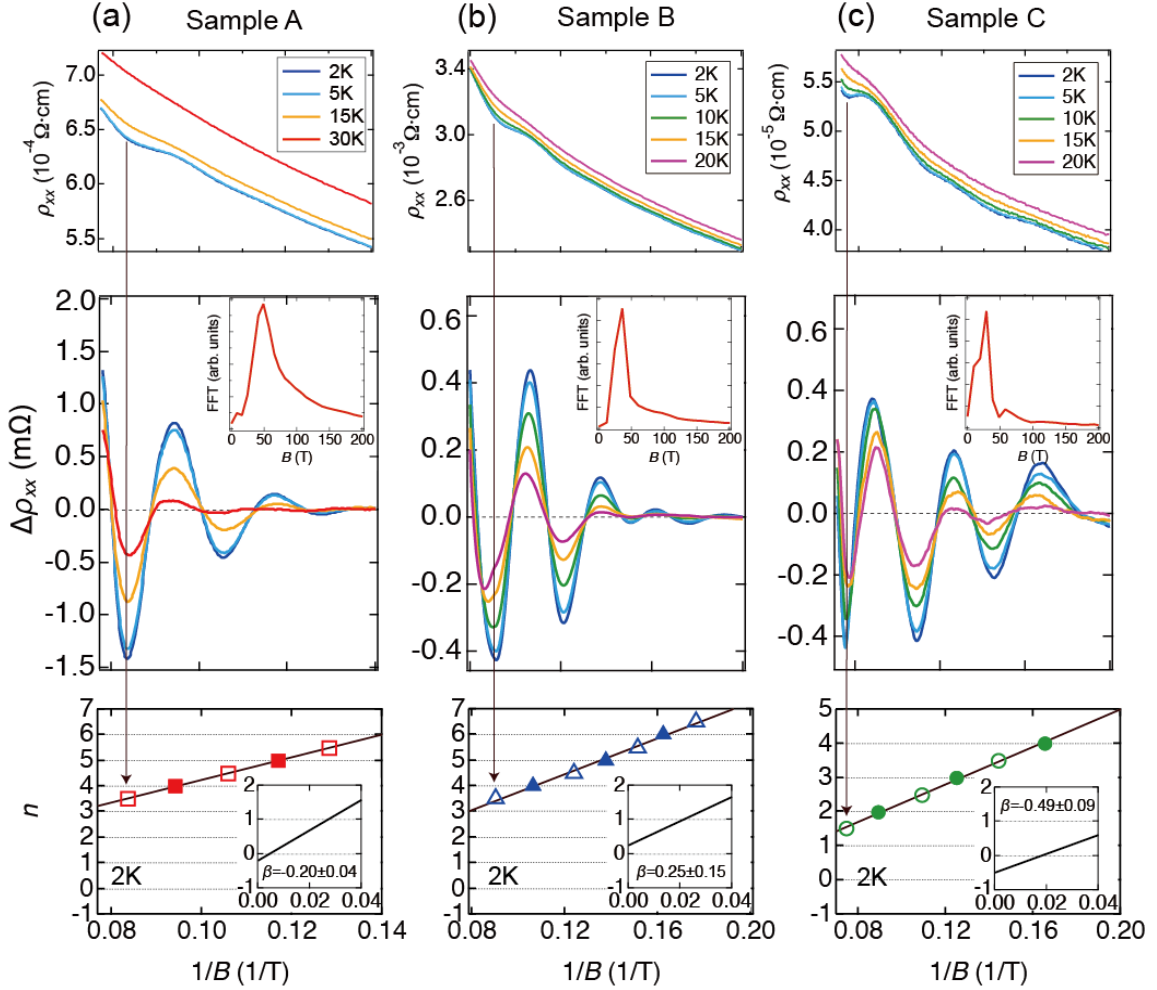

FIG. 4 (Color online) The Shubnikov-de Haas oscillations plotted as $\rho_{xx}$ (top) and $\Delta\rho_{xx}$ (middle) vs. the inverse magnetic field at some temperatures in Samples (a) A, (b) B, and (c) C, respectively. The $\Delta\rho_{xx}$ was derived by subtracting the background from $\rho_{xx}$. The background was given by smoothing with the Savitzky-Golay method. The insets of the middle figures represent the fast Fourier transform of oscillations to reveal the periodicities of the oscillations. The peak positions of FFT are 49, 36, and 29 T in Samples A, B, and C, respectively. The Landau-level fan diagrams are at the bottom of Figs. 4 (a) – (c). The vertical arrows indicate the positions of the corresponding oscillations. The open (close) red squares, blue triangles, and green circles mean the valleys (peaks) in $\Delta\rho_{xx}$ for Samples A, B, and C, respectively. The intercepts on $n$ axis shown in the insets of the fan diagrams correspond to $\beta$, giving the values of $|\beta| = 0.20 \pm 0.04$, $0.25 \pm 0.15$, and $0.49 \pm 0.09$ by line fitting.

The values of index *n* are chosen so that the intercept *β* on the vertical axis is close to zero as much as possible. *α* was estimated to be 45.4 ± 0.6, 35.0 ± 1.1, and 27.4 T ± 0.7 in Sample A, B and C, respectively, and their values are consistent with frequencies at the peak position in FFT spectra of Fig. 4. On the other hand, when |*β*| is close to 0.5 (or 0), the phase shift of the Berry phase is π (or 0). The intercept *β* in Samples A, B, and C were estimated to be -0.20 ± 0.04, 0.25 ± 0.15, and -0.49 ± 0.09, respectively. When their Berry phase is not π, rather intermediate values between 0 and π, it indicates that the electrical transport is not purely due to the TSS, but also due to contribution from the bulk states. Here, the |*β*| value of for Sample C is 0.49 ± 0.09, close to 0.5, indicating the Berry phase of π, which means that the oscillation comes from the helical TSS. However, since the error for the obtained *β* is relatively large, we need to check whether the SdH oscillations is purely due to the TSS or not by investigating the dependence of oscillations on the angle of magnetic field with respect to the sample surface, which distinguishes the dimensionality of the transport (as described later).

It is well known that an area of the cross section of the Fermi surface $S_F$ perpendicular to the applied magnetic field can be derived by Lifshitz-Onsager quantization rule using the gradient *α* in the Landau-level fan diagram in Fig. 4,

$$S_F = \frac{2\pi e}{\hbar}\alpha, \qquad (1)$$

where $e$ is the elementary charge and $\hbar$ is the Plank's constant divided by $2\pi$. The resultant values of $S_F$ in Samples A, B, and C are $4.2 \times 10^{-3}$, $3.3 \times 10^{-3}$, and $2.6 \times 10^{-3}$ $Å^{-2}$, respectively. Here, by assuming a circular Fermi surface, the relations of $S_F = \pi k_F^2$ ($k_F$ : Fermi wave vector) and $S_F = (2\pi)^2 n_{2D}$ ($n_{2D}$ : two-dimensional carrier density) are given. Therefore, $k_F$ and $n_{2D}$ of Samples A, B, and C were estimated to be $3.7 \times 10^{-2}$, $3.3 \times 10^{-2}$, $2.9 \times 10^{-2}$ $Å^{-1}$, and $1.1 \times 10^{12}$ (holes), $8.5 \times 10^{11}$ (holes), and $6.5 \times 10^{11}$ (electrons) cm$^{-2}$, respectively. These values of $n_{2D}$ are $\sim 10^3$ times smaller than those determined by the Hall measurements in Fig. 2(b). This is because the SdH oscillations at high magnetic fields detect preferentially carriers having higher mobility whereas the Hall measurements count all carriers including even low mobility in the bulk band. Inversely, this result means that all samples contain the bulk contribution in transport.

To analyze the characteristics of the transport further, the oscillating component in conductivity $\Delta\sigma_{xx}(= 1/\Delta\rho_{xx})$ at some temperatures was derived by subtracting the background from the data in Figs. 3(a) - (c), as shown in Figs. 5(a) - (c) for Samples A, B, and C at higher magnetic field region ($> \sim 8$ T). According to the Lifshitz−Kosevich (LK) theory [29], the temperature dependence of the amplitude in the SdH oscillation is described by

$$\frac{\Delta\sigma_{xx}(T)}{\Delta\sigma_{xx}(0)} = \frac{\lambda(T)}{\sinh(\lambda(T))}, \qquad (2)$$

with the thermal factor $\lambda(T) = 2\pi^2 m_{cyc} k_B T/\hbar eB$, where $m_{cyc}$ is the cyclotron mass and $k_B$ is the Boltzmann constant.

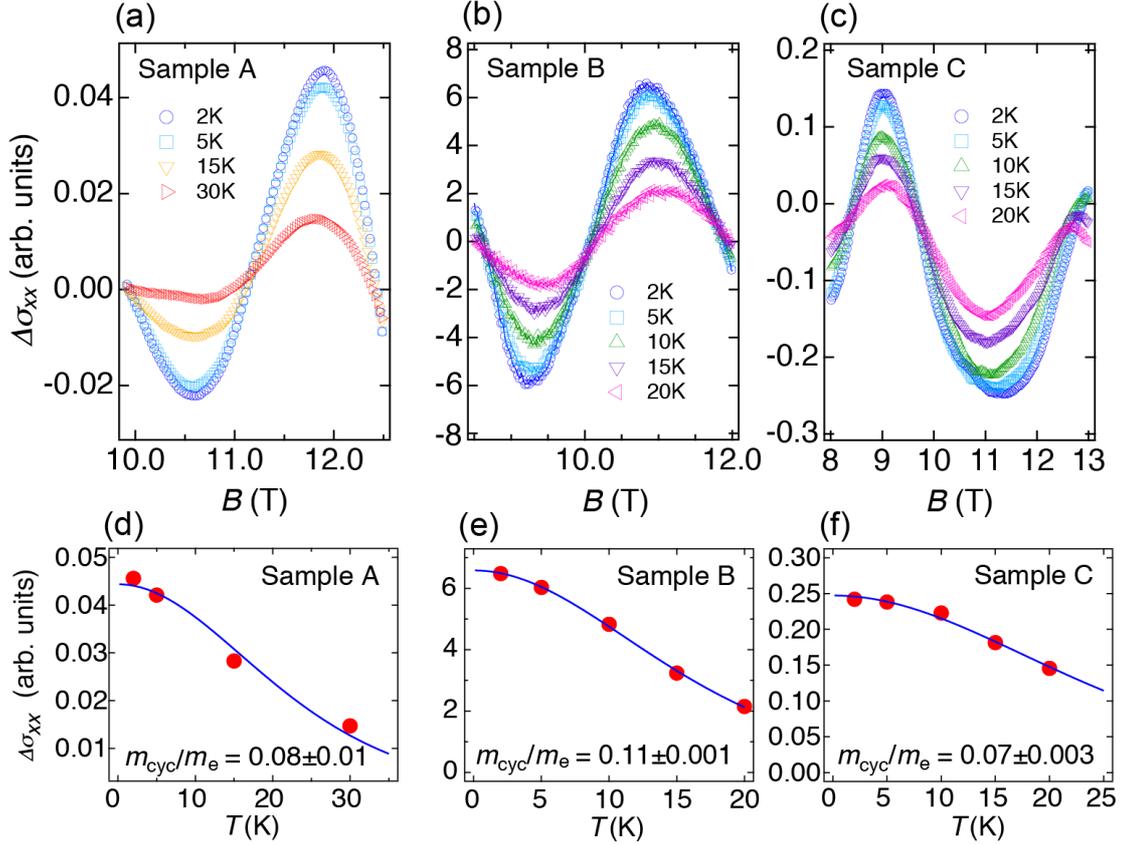

FIG. 5 (Color online) The oscillation in magnetoconductivity $\Delta\sigma_{xx}$ in the region of high magnetic fields at some temperatures for Samples (a) A, (b) B, and (c) C, respectively, taken from Fig. 3. (d) - (f). To estimate the cyclotron mass, the temperature dependences of amplitudes of the oscillation in $\Delta\sigma_{xx}$ are plotted (red circles), and are fitted by the standard Lifshitz-Kosevich (LK) theory (solid lines) for Samples (d) A, (e) B, and (f) C, respectively. The data are at peaks/valley in Sample A, B, and C at $B$ = 11.8, 11.1, and 11.2 T, respectively.

The values at peaks or valley in Figs. 5(a)-(c) were fitted with the LK theory, as shown in Figs. 5(d) - (f) for the respective samples. The resultant ratios of the cyclotron mass to the electron rest mass, $m_{cyc}/m_e$, of Samples A, B, and C are $0.08 \pm 0.01$, $0.110 \pm 0.001$, and $0.070 \pm 0.003$, respectively. All values are between the reported values of $m_{cyc}/m_e$ in $Sb_2Te_3$ (0.065 [30]) and $Bi_2Te_3$ (0.16 [31]). Then, the Fermi velocity can be estimated to be $v_F = \hbar k_F/m_{cyc} = 5.1 \times 10^5$, $3.5 \times 10^5$, and $4.8 \times 10^5$ m/s in Samples A, B, and C, respectively.

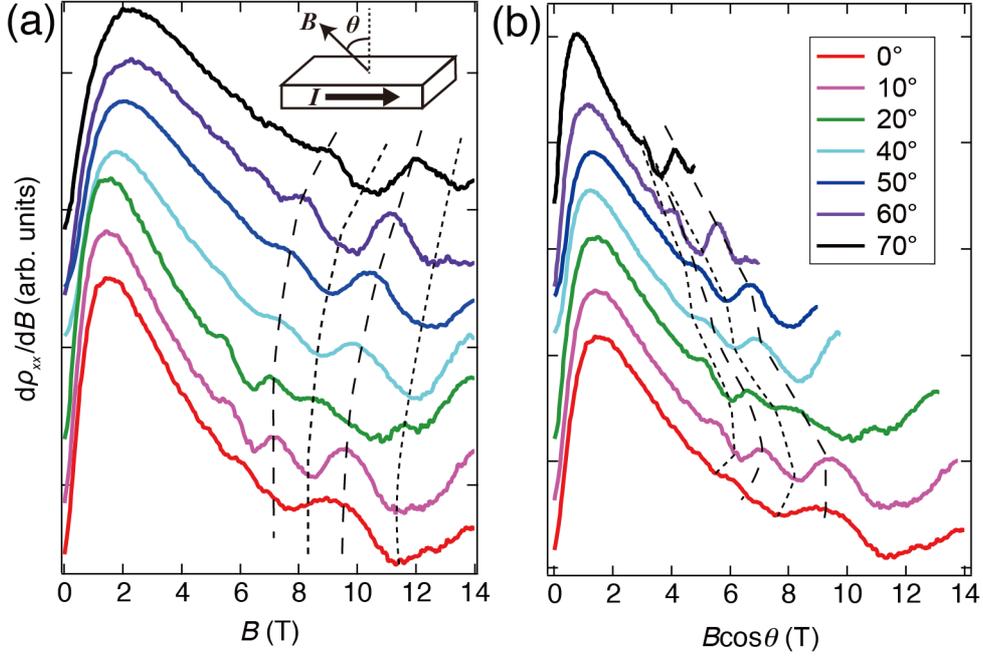

FIG. 6 (Color online) SdH oscillations of Sample C at 2 K at different angles $\theta$ of the magnetic field with respect to the surface-normal, plotted as a function of (a) $B$ and (b) $B\cos\theta$, respectively. The broken (dotted) lines trace peaks (valleys) in the oscillations. If the SdH oscillations have two (or three)-dimensional properties, the peaks and valleys should be at the same positions among different angles $\theta$ when plotted as a function of $B\cos\theta$ (or $B$). However, the data show neither two- nor three-dimensional.

The SdH oscillations of Sample C with different angles $\theta$ of magnetic field are shown in Figs. 6(a) and (b). Here, $\theta$ indicates the angle between the direction of magnetic field and the surface-normal direction as shown in the inset of Fig. 6(a). If the oscillation has three-dimensional characteristics, the positions of peaks and valleys do not depend on $\theta$ in Fig. 6(a). On the other hand, if the oscillation is two-dimensional case, it should depend only on the magnetic field component perpendicular to the surface, and thus the positions of peaks and valleys do not depend on $\theta$ in Fig. 6(b) where the data is plotted as a function of $B\cos\theta$. However, both Figs. 6 (a) and (b) show neither three- nor two-dimensional behaviors. This fact suggests that the transport channel in Sample C shows the surface-bulk coupling, and the dimensionality of the transport cannot be specified simply.

The lifetime of carriers contributing to the SdH oscillations can be estimated by the Dingle plot. According to the theory, the oscillation amplitude $\Delta\rho_{xx}$ in Fig. 4 is given by

$$\Delta\rho_{xx} = \frac{4\rho_0 \lambda e^{-\lambda_D}}{\sinh(\lambda)}, \tag{3}$$

where $\rho_0$ means the resistivity at zero magnetic field and $\lambda_D = 2\pi^2 m_{cyc} k_B T_D/\hbar eB$, respectively. $T_D$ is the Dingle temperature, $\hbar/2\pi k_B \tau$, where $\tau$ is the total scattering time. Figure 7(a) represent the plots of $\ln[\Delta\rho_{xx}\sinh(\lambda)/4\rho_0\lambda]$ vs. $1/B$ for Samples A, B, and C at 2 K. The $\tau$ can be estimated by the gradient of the fitting lines in Fig. 7(a), resulting in the values $\tau$ of 0.10, 0.37, and 0.67 ps for Samples A, B, and C, respectively.

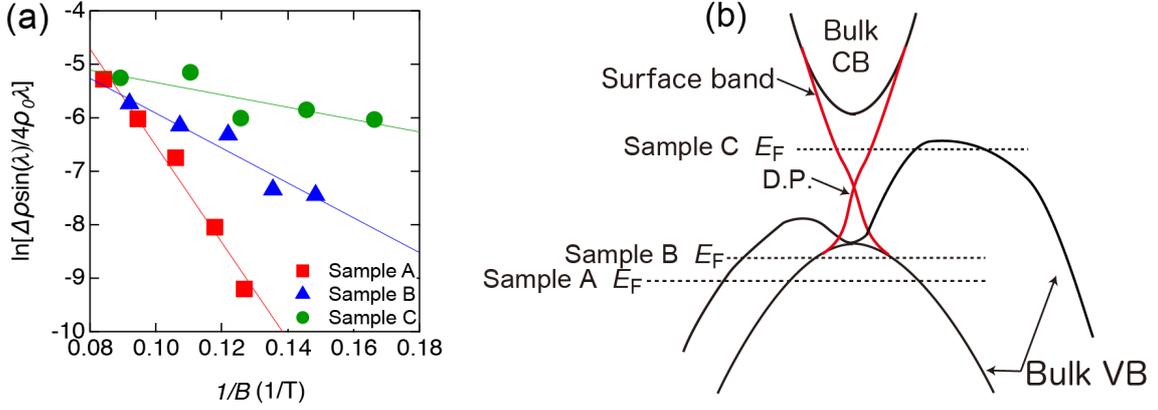

FIG. 7 (Color online) (a) The red squares, blue triangles, and green circles represent the Dingle plot of Samples A, B, and C at 2 K, respectively, and the solid lines are the fitting results. The slope of line gives the total scattering time $\tau$ as 0.10, 0.37, and 0.67 ps in Samples A, B, and C, respectively. (b) A schematic picture of the band dispersion with Fermi level ($E_F$) estimated from the results of Hall effect, the Berry phase and SdH oscillations.

The mean-free path $l = v_F \tau$ and the mobility of carriers $\mu = e\tau/m_{cyc}$ of Samples A, B, and C are then estimated to be 53, 130, 320 nm, and 2000, 6000, and 17,000 cm$^2$/V/s, respectively. It should be noted that the mobility becomes high for Sample C compared with Samples A and B, due to the surface-bulk coupling. This character cannot be induced only by the bulk bands. If this enhancement of mobility in Sample C is solely due to the difference of the carrier type, such enhancement should be mainly originated from the low effective mass of carriers because crystal qualities of samples with Sample B: *p*-type and Sample C: *n*-type do not differ from each other significantly. However, in our case the effective mass does not change much among samples: 0.08, 0.11 and 0.07 in Sample A, B and C, respectively. This means that the enhanced mobility in Sample C is mainly due to decrement of the scattering. It is suggested that this decrement is enabled by the existence of the surface band which prohibits the backscattering.

Table 1.

|  | Sample A | Sample B | Sample C |
| --- | --- | --- | --- |
| Concentration $x$ in $(Bi_xSb_{1-x})_2Te_3$ | 0.29 | 0.34 | 0.42 |
| Type of carriers | holes | holes | electrons |
| 2D carrier density (cm$^{-2}$) | $1.1 \times 10^{12}$ | $8.5 \times 10^{11}$ | $6.5 \times 10^{11}$ |
| Cyclotron mass ($m_{cyc}/m_e$) | $0.08 \pm 0.01$ | $0.110 \pm 0.001$ | $0.070 \pm 0.003$ |
| Total scattering time $\tau$ (p sec) | 0.10 | 0.37 | 0.67 |
| Mean-free path (nm) | 53 | 130 | 320 |
| Fermi wave number $k_F$ (Å$^{-1}$) | $3.7 \times 10^{-2}$ | $3.3 \times 10^{-2}$ | $2.9 \times 10^{-2}$ |
| Fermi surface area $S_F$ (Å$^{-2}$) | $4.2 \times 10^{-3}$ | $3.3 \times 10^{-3}$ | $2.6 \times 10^{-3}$ |
| Intercept of the fan diagram | $0.20 \pm 0.04$ | $0.25 \pm 0.15$ | $0.49 \pm 0.09$ |
| Fermi velocity (m/s) | $5.1 \times 10^5$ | $3.5 \times 10^5$ | $4.8 \times 10^5$ |
| Mobility (cm$^2$/V/s) | 2,000 | 6,000 | 17,000 |

The physical parameters of Samples A, B, and C obtained in the present study. The cyclotron mass $m_{cyc}$ and the total scattering time $\tau$ were derived by the LK-theory and the Dingle plot. The mean-free path $l = v_F \tau$, the cross section area of the Fermi surface $S_F = \frac{2\pi e}{\hbar} \alpha$ ($\alpha$ was given by the gradient of fitted line in the fan diagram.), and the Fermi wave vector $k_F = \sqrt{\frac{S_F}{\pi}}$ were calculated from the experimental data.

The estimated parameters obtained in this study are summarized in Table 1. The Fermi-level positions of Samples A, B and C can be roughly indicated as shown in Fig. 7(b) by considering the carrier density, the carrier type, and the Berry phase. Whereas the Fermi levels of Sample A and B lie in the bulk valence band, the Fermi level of Sample C lies across the bulk valence band as well as TSS band above the Dirac point since the carrier type is $n$-type and SdH oscillations show the surface-bulk coupling transport. This band configuration was confirmed by ARPES measurements and ab initio calculations in Figs.1 (b) and (d) of Ref. [11], which show those of samples with Bi/Sb content of 50/50 ($x$=0.5) and 25/75 ($x$=0.25). In our Sample C case Bi/Sb = 42/58 ($x$=0.42), and thus the band state seems to be intermediate between them. From these figures, we can see that the Fermi level cuts across both the surface conduction band and the bulk valence band. Since the Fermi level just touches the top of the bulk valence band, the conduction is dominated by the surface-state electrons over the bulk holes, resulting in $n$-type.

In summary, we observed SdH oscillations in the tape-cleaved flakes of $(Bi_xSb_{1-x})_2Te_3$, with $x$ = 0.29 (Sample A), 0.34 (Sample B), and 0.42 (Sample C). The Hall-effect measurements indicated that the carrier type was $n$-type in Sample C whereas the others were $p$-type. The intercept at $1/B = 0$ in the fan diagram for Sample C indicates the Berry phase of $\pi$ and a high carrier mobility (17,000 cm$^2$/V/s), while Samples A and B have the Berry phase of nearly zero and lower mobilities. While the transport at Samples A and B is dominated by the bulk states, Sample C shows surface-bulk coupling transport, deduced by the analysis of the transport dimensionality in SdH oscillations. The surface-bulk coupling suppresses the carrier scattering and makes the mobility higher than the bulk one. These represent that electrical properties can be changed largely by tuning the concentration $x$ in $(Bi_xSb_{1-x})_2Te_3$.

**Acknowledgments**

This work was partially supported by a Grant-in-Aid for Scientific Research (A) (KAKENHI No. JP16H02108, 26247064), a Grant-in-Aid for Young Scientists (B) (KAKENHI No. 26870086), Innovative Areas "Topological Materials Science" (KAKENHI No. JP16H00983, JP15K21717), "Molecular Architectonics" (KAKENHI No. 25110010) from Japan Society for the Promotion of Science, the bilateral collaboration program between RFBR (Russia) and JSPS (Japan), the Saint Petersburg State University project No. 15.61.202.2015. and the Russian Science Foundation project No. 17-12-01047. KS was financially supported by Grant-in-Aid for JSPS Fellows (No.16J03874).